\documentclass[journal]{vgtc}                

\usepackage{mathptmx}
\usepackage{graphicx}
\usepackage{times}
\usepackage{amsmath}

\usepackage{url}
\usepackage{subcaption}
\let\ifpdf\relax
\usepackage{hyperref}

\onlineid{0}

\vgtccategory{Research}

\vgtcinsertpkg

\title{Segmentation of blood vessels in retinal fundus images}

\manuscriptnote{}

\author{Michiel Straat and Jorrit Oosterhof}
\authorfooter{
\item Michiel Straat is a first year Computing Science master student at the University of Groningen.
\item Jorrit Oosterhof is a first year Computing Science master student at the University of Groningen.
}

\shortauthortitle{Biv \MakeLowercase{\textit{et al.}}: Global Illumination for Fun and Profit}

\abstract{In recent years, several automatic segmentation methods have been proposed  for blood vessels in retinal fundus images, ranging from using cheap and fast trainable filters \cite{BCOSFIRE} to complicated neural networks and even deep learning \cite{Fraz} \cite{Krawiec} \cite{Prokofyeva}.\\
\hspace*{10pt}One example of a filted-based segmentation method is B-COSFIRE \cite{BCOSFIRE}. In this approach the image filter is trained with example prototype patterns, to which the filter becomes selective by finding points in a Difference of Gaussian response on circles around the center with large intensity variation.\\
\hspace*{10pt}In this paper we discuss and evaluate several of these vessel segmentation methods. We take a closer look at B-COSFIRE and study the performance of B-COSFIRE on the recently published IOSTAR dataset \cite{IOSTAR} by experiments and we examine how the parameter values affect the performance. In the experiment we manage to reach a segmentation accuracy of 0.9419. \\
\hspace*{10pt}Based on our findings we discuss when B-COSFIRE is the preferred method to use and in which circumstances it could be beneficial to use a more (computationally) complex segmentation method. We also shortly discuss areas beyond blood vessel segmentation where these methods can be used to segment elongated structures, such as rivers in satellite images or nerves of a leaf.
} 


\CCScatlist{ 
  \CCScat{K.6.1}{Management of Computing and Information Systems}%
{Project and People Management}{Life Cycle};
  \CCScat{K.7.m}{The Computing Profession}{Miscellaneous}{Ethics}
}

\begin{document}


\firstsection{Introduction}

\maketitle


The inspection of the blood vessel tree in the fundus, which is the interior surface of the eye opposite to the lens, is important in the determination of various cardiovascular diseases. This can be done manually by ophthalmoscopy, which is an effective method of analysing the retina. However, it has been suggested that using fundus photographs is more reliable than ophthalmoscopy \cite{BCOSFIRE}. Additionally, these images can be used for automatic identification of the blood vessels, which can be a difficult task due to obstacles such as low contrast with the background, narrow blood vessels and various blood vessel anomalies. A segmentation method with high accuracy can serve as a significant aid in diagnosing cardiovascular diseases, as it highlights the blood vessel tree in the fundus.

In recent years, several segmentation methods have been proposed for the automatic segmentation of blood vessels, ranging from using cheap and fast trainable filters \cite{BCOSFIRE} to complicated neural networks and even deep learning \cite{Fraz} \cite{Krawiec} \cite{Prokofyeva}.

One example of a filted-based segmentation method is Bar-Combination Of Shifted Filter Responses, in short B-COSFIRE \cite{BCOSFIRE}. In this approach we train the image filter with example prototype patterns, to which the filter becomes selective by finding points in a Difference of Gaussian response on circles around the center with large intensity variation. Azzopardi \textit{et al}. used two prototype patterns for the segmentation: One for bars and one for bar-endings. The filter achieves rotation invariance by rotating the points of interest for the prototype patterns, yielding a filter that is selective for these vessel orientations as well.

In \autoref{sec:BCOSFIRE} we discuss the theory behind the filter-based method B-COSFIRE in more detail. We then discuss supervised vessel segmentation methods that are based on machine learning in \autoref{sec:MLmethods}. In \autoref{sec:experiments} we describe our set-up for experiments with B-COSFIRE on a recent dataset that contains retinal images acquired with a camera based on Scanning Laser Ophtalmoscopy (SLO) technology and we discuss the results in \autoref{sec:results}. Based on the study and the results of the experiments we then discuss the advantages- and disadvantages of the discussed methods.

\section{B-COSFIRE}\label{sec:BCOSFIRE}
In this section we discuss B-COSFIRE in more detail. The method is based on the \textbf{C}ombination \textbf{O}f \textbf{S}hifted \textbf{FI}lter \textbf{RE}sponses, which is a trainable filter used for interest point detection, as described in \cite{COSFIRE}. The filter is trained in a training stage in which it is configured to be selective to specific prototype patterns.

The B-COSFIRE method is a specific case of COSFIRE, in which the B stands for "bar". The filter is trained using vessel-like prototype patterns, which allows it to be selective for such structures. Azzopardi \textit{et al}. proposed two B-COSFIRE filters: The \textit{symmetric} filter, which is suitable for bars, and the \textit{asymmetric} filter, suitable for bar endings. In the following, we briefly discuss the method as outlined in \cite{BCOSFIRE}.

\begin{figure}[htb]
  \centering
  \begin{subfigure}{0.22\textwidth}
  	\includegraphics[width=\textwidth]{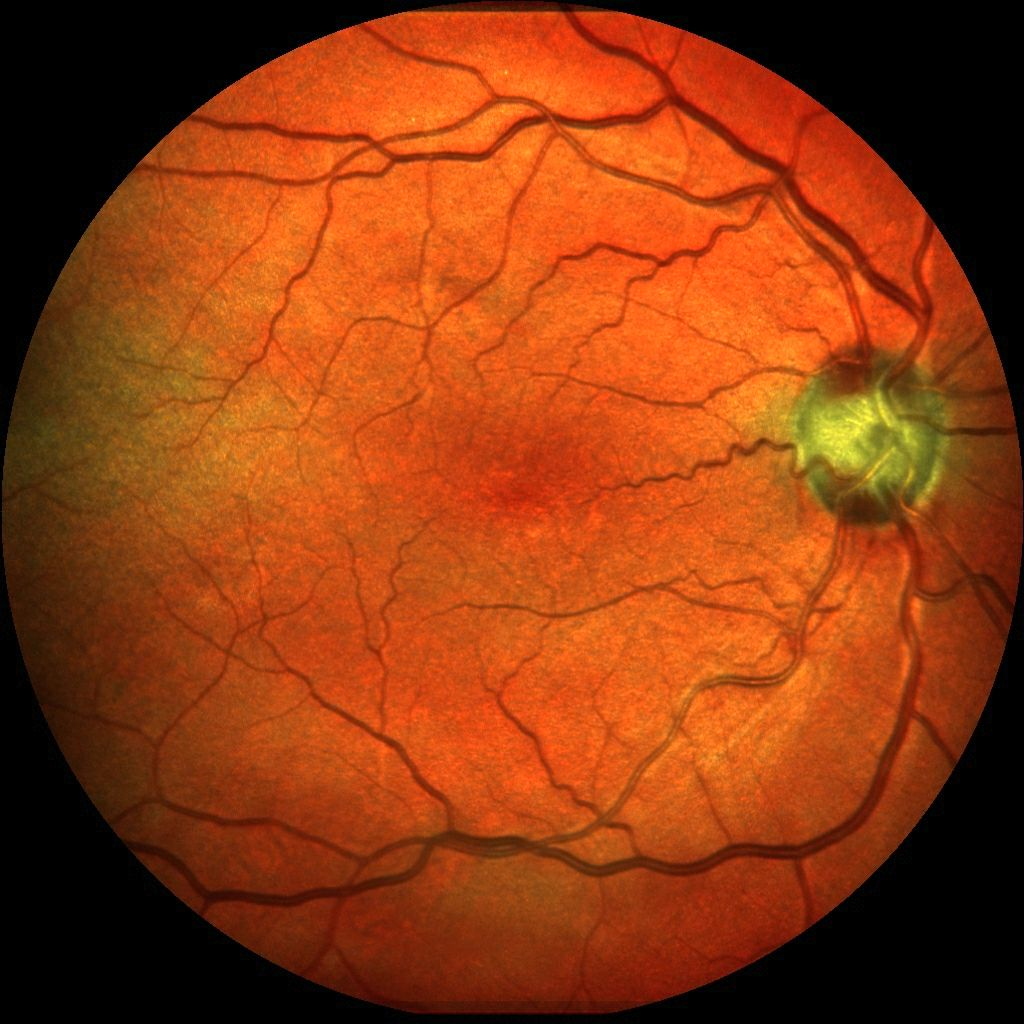}
  	\caption{Retinal fundus image}
  	\label{fig:IOSTAR2}
  \end{subfigure}
  \begin{subfigure}{0.22\textwidth}
  	\includegraphics[width=\textwidth]{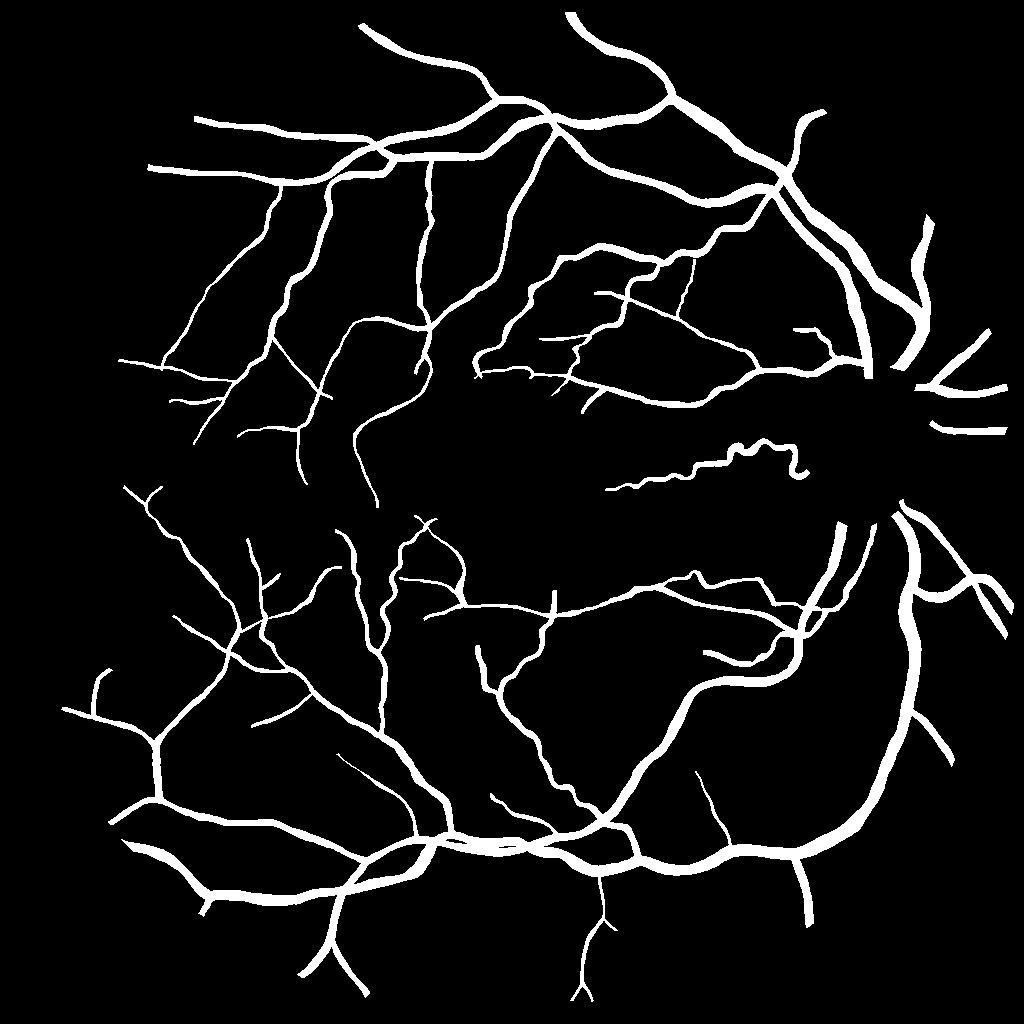}
  	\caption{Corresponding vessel segmentation.}
  \end{subfigure}
  \caption{A typical vessel tree extracted from a retinal image, in which each pixel is labeled either "vessel" (white) or "non-vessel" (black).}
  \label{fig:treeExample}
\end{figure}

The B-COSFIRE filtering method proceeds in a number of stages, which are roughly divided in training each filter with a prototype pattern in order for the filter to become selective for its prototype pattern, pre-processing the retinal input image in order to enhance contrast of the vessels and the actual filtering, which yields the final output response. The response is thresholded to obtain a binary image, in which for each pixel in the response a gray value above the threshold is shown as a white pixel (1) indicating "vessel", and a gray value below the threshold is shown as black indicating "non-vessel". This yields the typical vessel tree image shown in \autoref{fig:treeExample}.

In the next sections we discuss the stages mentioned above in more detail.

\subsection{Training a filter using a prototype pattern}
Difference of Gaussian (DoG) filters are used to detect changes in intensity.

\begin{equation} \label{eq:DoGFilter}
DoG_\sigma(x,y) = \frac{1}{2\pi \sigma^2} \exp\left(-\frac{x^2 + y^2}{2\sigma^2}\right) - \frac{1}{2\pi (0.5\sigma)^2} \exp\left(-\frac{x^2 + y^2}{2(0.5\sigma)^2}\right)\ .
\end{equation}

The response by applying a DoG filter as in \autoref{eq:DoGFilter} to an image $f$ is defined as the convolution of $f$ with the filter kernel in \autoref{eq:DoGFilter}:
\begin{equation} \label{eq:convolutionDoG}
	c_\sigma(x,y) = |f \star DoG_\sigma|^+\ .
\end{equation}

The selectivity of a B-COSFIRE filter is automatically configured by presenting the filter with a prototype pattern. Then, a number of concentric circles are put around the filter's center of support and the points on these circles with a high intensity variation are determined, called the points of interests. One such point is described by a triple $(\sigma, \rho, \phi)$, where $\sigma$ is the standard deviation of the DoG filter that contributed this point, $\rho$ is the radius of the circle that this point lies on and $\phi$ is the angle with respect to the center of support. The training of one prototype pattern yields a set of triples $S = \{(\sigma_i, \rho_i, \phi_i)\, i=1,...,n\}$ describing $n$ points of interest. 

\begin{figure}[htb]
  \centering
  \begin{subfigure}{0.22\textwidth}
  	\includegraphics[width=\textwidth,height=\textwidth]{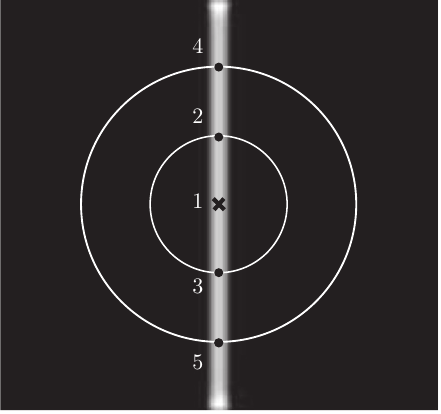}
  	\caption{Symmetric filter}
  	\label{fig:symmetricBar}
  \end{subfigure}
  \begin{subfigure}{0.22\textwidth}
  	\includegraphics[width=\textwidth,height=\textwidth]{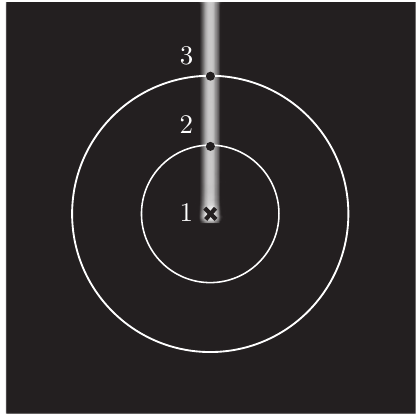}
  	\caption{Asymmetric filter}
  	\label{fig:asymmetricBar}
  \end{subfigure}  
  \caption{(a): A DoG filter applied to a straight full vertical bar pattern. After applying the DoG filter four local maxima are detected on the two concentric circles surrounding the center point, i.e. $|S|=5$ (including centerpoint). These correspond to the positions of the highest intensity variations around the center point. (b): A DoG filter applied to a straight half vertical bar pattern. Two local maxima are found on the circles, i.e. $|S|=3$ (including centerpoint). Images taken from \cite{BCOSFIRE}.}
  \label{fig:config}
\end{figure}

In Figure \ref{fig:symmetricBar} we see an example of how selectivity is obtained for a vertical bar pattern. The center point is surrounded by two circles. On each circle two local maxima in the DoG response are found, where two are above the center point and the other two below the center point. For this reason we call this the \textit{symmetric filter}. In Figure \ref{fig:asymmetricBar} an example is shown of the configuration of a bar-ending. In this case two points above the center point are marked as local maxima in the DoG response. This filter we call \textit{asymmetric}.

As the selectivity of the filter directly depends on the specified prototype patterns, by training with the two prototype patterns in Figure \ref{fig:config} the filter is only selective for vertical bars and bar-endings. To achieve selectivity also for other vessel orientations, we could train bar prototype patterns with more orientations. A more easy and efficient way to achieve rotation invariance is to transform the set of points $S$ that were found for the vertical bar patterns to a new set:

\begin{equation}
R_\psi(S) = \{ (\sigma_i,\rho_i,\phi_i + \psi) | \forall(\sigma_i,\rho_i,\phi_i) \in S \}\, ,
\end{equation}
where $\psi$ is the angle of the orientation of the bar for which the set $R_\psi(S)$ is selective. In the symmetric case, by taking $\psi=\frac{\pi}{12}, \frac{\pi}{6},..., \frac{11\pi}{12}$, we obtain a filter that, including the original vertical bar orientation, becomes selective for 12 bar orientations. In the asymmetric case we must naturally consider 24 orientations, as in this case a rotation by $\pi$ gives rise to a different orientation due to the asymmetry.

\subsection{Pre-processing the retinal image} \label{sec:preprocess}
\begin{figure}[htb]
  \centering
  \begin{subfigure}{0.22\textwidth}
  	\includegraphics[width=\textwidth]{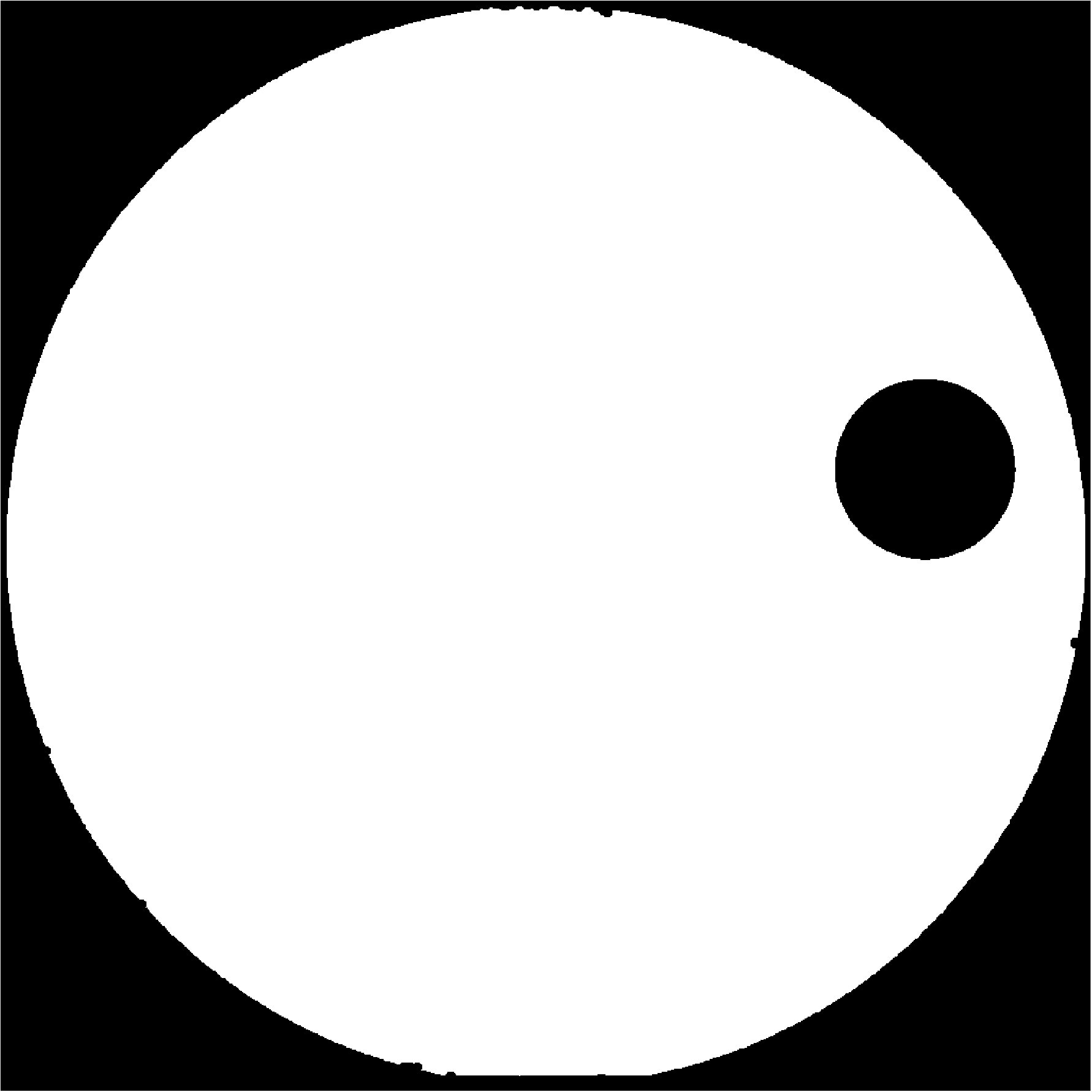}
  	\caption{Mask}
  	\label{fig:IOSTAR2Mask}
  \end{subfigure}
  \begin{subfigure}{0.22\textwidth}
  	\includegraphics[width=\textwidth]{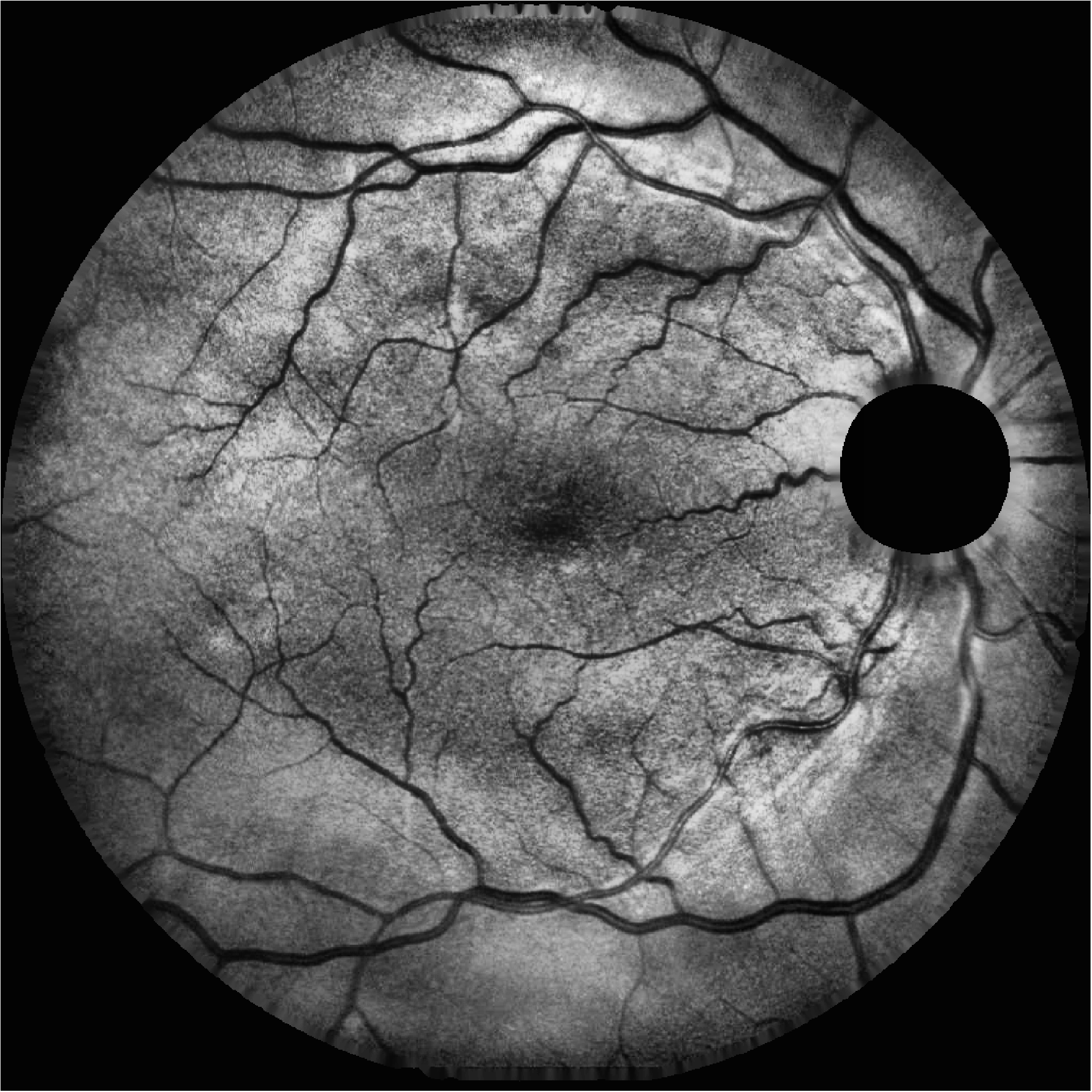}
  	\caption{Green channel}
  	\label{fig:IOSTAR2GreenPre}
  \end{subfigure}  
  \caption{(a): The mask that is used to segment the retina including an indication of the optic disc. (b): The resulting green channel of a retinal image after pre-processing is applied as described in \autoref{sec:preprocess}.}
  \label{fig:preprocess}
\end{figure}

We follow the pre-processing steps as discussed by Azzopardi \textit{et al}. Previous works have shown that the green channel of RGB images defines the contrast between vessels and the background better than the red channel, which has low contrast, or the blue channel, which shows a small dynamic range \cite{staal:2004-855}\cite{1677726}\cite{1677727}\cite{4336179}\cite{niemeijer:2004-866}. Therefore, we only use the green channel for the segmentation. The Field Of View (FOV) masks are provided with the dataset, and we smooth the borders around the FOV to ensure there will be no false positives in these areas because of the high contrast. The dataset IOSTAR also comes with masks that indicate the Optic Disc (OD) in the retina. We use these masks instead of the normal FOV masks to ensure that segmentation will not take place inside the OD, since the ground truth images do not have a specified segmentation inside the OD area as well, which gives rise to a better comparison. In the last step the contrast-limited adaptive histogram equalization is applied. The output of the pre-processing stage for the retinal image of \autoref{fig:IOSTAR2} is given in \autoref{fig:IOSTAR2GreenPre}.

\subsection{Filter application}
We have shown how a B-COSFIRE filter can achieve selectivity for specific bar patterns, by selecting points around center of support of the filter with a high intensity variation. We have also shown how from there, rotation-invariance can be easily achieved by rotating the points of interest by angles $\psi$. We have shown how this works for two prototype patterns that give rise to a symmetric- and an asymmetric bar filter. 
In order to achieve a more tolerant filter and also a ranking of the points of interest according to their importance such that the points more closely to the center point get a higher weight, the DoG responses are blurred with a Gaussian $G_{\sigma'}(x',y')$. Because closer points need to get more weight, the standard deviation of the Gaussian is defined in terms of $\rho_i$, the radius of the circle on which the point of interest lies:

\begin{equation} \label{eq:sigmaGaussian}
	\sigma' = \sigma_0' + \alpha \rho_i\ ,
\end{equation}
where $\sigma_0$ is a constant for the base standard deviation and $\alpha$ is the rate at which the standard deviation increases. Therefore, the higher $\alpha$, the more tolerant the filter becomes. The DoG responses are moved to the center of the filter support by shifting with $(\Delta x_i = -\rho_i \cos \phi_i, \Delta y_i = -\rho_i \sin \phi_i)$. The $i$th blurred and shifted DoG response then becomes:
\begin{equation} \label{eq:blurDoG}
 S_{\sigma_i,\rho_i,\phi_i} (x,y) =  \max_{x',y'} \{c_{\sigma_i} ( x - \Delta x_i - x', y - \Delta y_i - y')G_{\sigma'}(x',y') \}\ ,
\end{equation}
where we consider a neighborhood around the points of interest of $-3\sigma' \leq x',y' \leq 3\sigma'$. As we can see, the maximum weighted response is the value we assign to $s_{\sigma_i,\rho_i,\phi_i}(x,y)$.
Now at each pixel of the input image, each trained B-COSFIRE filter is applied by multiplying the now obtained blurred and shifted DoG responses in the set $S$ in the following way:

\begin{multline} \label{eq:filterApplication}
r(x,y) = \left( \prod_{i=1}^{|S|} ( S_{\sigma_i,\rho_i,\phi_i}(x,y) )^{\omega_i} \right)^{ 1/\sum_{i=1}^{|S|} \omega_i } \\ 
\omega_i = \exp^{-\frac{\rho^2}{2\hat{\sigma}^2}},\ \hat{\sigma} = \frac{1}{3} \max_{i \in \{1...|S|\}} \{ \rho_i \}\, ,
\end{multline}
where we obtain $r_s(x,y)$ for the symmetric filter response and $r_a(x,y)$ for the asymmetric filter response. As a small sidenote, \autoref{eq:filterApplication} shows exactly why the filter is called COSFIRE: The response is a \textit{combination}, which refers to the product, of \textit{shifted} filter responses, which refers to the shifts of the filter responses by $(\Delta x_i, \Delta y_i)$ to the center of the B-COSFIRE filter support.

The sum of the symmetric- and the asymmetric filter $r_{as}(x,y) = r_s(x,y) + r_a(x,y)$ is then the total response of the filtering process. To get a binary image in which each pixel is classified as either vessel or non-vessel, we threshold the filter response $r_{as}(x,y)$ with threshold $T$, where pixel values of $r_{as}(x,y)$ above $T$ are considered as vessel pixels, i.e.:

\begin{equation} \label{eq:thresholding}
	g(x,y) = 
\begin{cases}
	1,&\text{if } r_{as}(x,y) > T\\ 
	0,&\text{if } r_{as}(x,y) \leq T
\end{cases}
\end{equation}

\section{Alternative methods} \label{sec:MLmethods}
The algorithms for segmenting blood vessels in retinal fundus images can be divided in supervised and unsupervised methods. B-COSFIRE is an example of an unsupervised approach to vessel segmentation. Supervised methods first train on a set of example images with their ground truth segmentation provided. \cite{Fraz2012407} discusses several blood vessel detection methods that were published by 2012. In this section, we discuss a few of these methods and compare them with each other. \autoref{tbl:DRIVEResults}, \autoref{tbl:STAREResults} and \autoref{tbl:CHASEResults} show some performance metrics of the methods we discuss.


\subsection{Deep Neural Networks}
One can use deep neural networks to extract blood vessels from a fundus image. Deep neural networks are neural networks with many (hidden) layers. Liskowski \textit{et al}. \cite{Krawiec} use deep learning to identify blood vessels in fundus images. The deep learning algorithm is the error back-propagation algorithm extended with dropout \cite{Srivastava:2014:DSW:2627435.2670313}. The error back-propagation algorithm is a common algorithm in the field of neural networks. With dropout, during training a percentage of units is temporarily disabled, introducing an extra challenge for the training process.

Liskowski \textit{et al}. use the DRIVE \cite{staal:2004-855}\cite{DRIVELink}, STARE \cite{Hoover00locatingblood}\cite{STARELink} and CHASE \cite{CHASELink} datasets to verify their deep learning method. To classify pixels as being either \textit{vessel} or \textit{non-vessel}, a patch of $m \times m$ pixels is used, centred at the pixel. The neural network is fed with a triple of such patches. Each of those patches is at the same location, but each patch is from one of the three RGB channels. As a result, this method uses all three RGB channels to determine whether a pixel belongs to a blood vessel or not. This is in contrast to the filter-based method B-COSFIRE, which only uses the green channel as if it were a grey scale image \cite{BCOSFIRE}, for reasons discussed in \autoref{sec:preprocess}.

For their experiments, Liskowski \textit{et al}. use multiple configurations. \autoref{tbl:DRIVEResults} and \autoref{tbl:STAREResults} show metrics for the two configurations they found best. Furthermore, Liskowski \textit{et al}. used structured prediction, an approach which also uses the neighbourhood of a pixel for the classification process. Separate results for the two best configurations using structured prediction are also shown in \autoref{tbl:DRIVEResults} and \autoref{tbl:STAREResults}. Please refer to \cite{Krawiec} for a detailed description of the configurations with and without structured prediction.

\subsection{Ensemble classification-based approach}
Fraz \textit{et al}. \cite{Fraz} use an ensemble classifier of boosted and bagged decision trees. In short, they use multiple classifiers to classify the pixels and take a majority vote to determine whether a pixel is a \textit{vessel} or \textit{non-vessel}. Similar to the B-COSFIRE method, this method uses the green channel of the RGB image for the blood vessel segmentation. For experimenting, Fraz \textit{et al}. use the DRIVE, STARE and CHASE datasets. The classifier is trained using a randomly selected subset of pixels for each image.

\subsection{Fully Connected Conditional Random Field Model}
Orlando \textit{et al}. \cite{Prokofyeva} use conditional random fields (CRFs). In contrast to normal classifiers, CRFs do not classify `objects' purely based on the object. Note that word `object' here can mean anything, like individual pixels or what activity belongs to an image. For example, for classifying that an image, from a sequence of snapshots of someone's life, displays an activity, it can be useful to know which activity was classified before.

CRFs work with graphs. Orlando \textit{et al}. use a fully connected CRF (FC-CRF) where each pixel, being a node in the graph of the CRF, is linked to every other pixel. This has the advantage that the method can take long-range interactions between pixels into account, instead of only neighbouring information, which results in an improvement of the segmentation accuracy of the method \cite{Prokofyeva}. Similar to the B-COSFIRE method, this method only uses the green channel of the RGB image for the blood vessel segmentation.

\begin{table}
	\centering
	\begin{tabular}{lcccc}
		\hline
		Method & AUC & Acc & Se & Sp\\
		\hline
		B-COSFIRE & 0.9614 & 0.9442 & 0.7655 & 0.9704\\
		\hline
		\multicolumn{5}{l}{Liskowski \textit{et al}. \cite{Krawiec}}\\
		Balanced & 0.9738 & 0.9230 & 0.9160 & 0.9241\\
		No-Pool & 0.9720 & 0.9495 & 0.7763 & 0.9768\\
		Balanced SP & 0.9788 & 0.9530 & 0.8149 & 0.9749\\
		No-Pool SP & 0.9790 & 0.9535 & 0.7811 & 0.9807\\
		\hline
		Fraz \textit{et al}. \cite{Fraz} & 0.9747 & 0.9480 & 0.7406 & 0.9807\\
		Orlando \textit{et al}. \cite{Prokofyeva} & - & - & 0.7897 & 0.9684\\
	\end{tabular}
	\caption{Results on the DRIVE dataset}
	\label{tbl:DRIVEResults}
\end{table}

\begin{table}
	\centering
	\begin{tabular}{lcccc}
		\hline
		Method & AUC & Acc & Se & Sp\\
		\hline
		B-COSFIRE & 0.9563 & 0.9497 & 0.7716 & 0.9701\\
		\hline
		\multicolumn{5}{l}{Liskowski \textit{et al}. \cite{Krawiec}}\\
		Balanced & 0.9820 & 0.9309 & 0.9307 & 0.9304\\
		No-Pool & 0.9785 & 0.9566 & 0.7867 & 0.9754\\
		Balanced SP & 0.9928 & 0.9700 & 0.9075 & 0.9771\\
		No-Pool SP & 0.9928 & 0.9729 & 0.8554 & 0.9862\\
		\hline
		Fraz \textit{et al}. \cite{Fraz} & 0.9768 & 0.9534 & 0.7548 & 0.9763\\
		Orlando \textit{et al}. \cite{Prokofyeva} & - & - & 0.7680 & 0.9738\\
	\end{tabular}
	\caption{Results on the STARE dataset}
	\label{tbl:STAREResults}
\end{table}

\begin{table}
	\centering
	\begin{tabular}{lcccc}
		\hline
		Method & AUC & Acc & Se & Sp\\
		\hline
		B-COSFIRE & 0.9487 & 0.9387 & 0.7585 & 0.9587\\
		\hline
		\multicolumn{5}{l}{Liskowski \textit{et al}. \cite{Krawiec}}\\
		\multicolumn{5}{c}{Using DRIVE for training}\\
		No-Pool & 0.9646 & 0.9473 & 0.7158 & 0.9810\\
		No-Pool SP & 0.9710 & 0.9515 & 0.7520 & 0.9806\\
		\multicolumn{5}{c}{Using STARE for training}\\
		No-Pool & 0.9543 & 0.9525 & 0.7091 & 0.9791\\
		No-Pool SP & 0.9880 & 0.9696 & 0.8145 & 0.9866\\
		\hline
		Fraz \textit{et al}. \cite{Fraz} & 0.9712 & 0.9469 & 0.7224 & 0.9711\\
		Orlando \textit{et al}. \cite{Prokofyeva} & - & - & 0.7277 & 0.9712\\
	\end{tabular}
	\caption{Results on the CHASE dataset}
	\label{tbl:CHASEResults}
\end{table}

\section{Experiments} \label{sec:experiments}
To study the performance of the filter-based B-COSFIRE method discussed in \autoref{sec:BCOSFIRE}, we experiment with the method on a recent retinal image dataset. In this section we report on the parameters that we use and how we evaluate the resulting segmentation.

\subsection{Training parameters}
A major advantage of the B-COSFIRE method is that the parameters are quite intuitive, so one can estimate a good set of parameter values for $(\sigma, \rho, \sigma_0, \alpha)$. In any case, the parameters must be carefully chosen, as the images of different retinal image dataset have varying properties. The parameters that are to be determined are:

\begin{itemize}
	\item $\sigma$: The standard deviation of the outer Gaussian function in the DoG filter. Intuitively, the higher the detail in the image, i.e., the higher the resolution, the greater $\sigma$ needs to be to still detect blood vessels with high accuracy, since they are made up of more pixels in high resolution images. 
	
%
	
	\item $\rho$: The set of radii of the concentric circles surrounding the filter's center of support.
	\item $\sigma_0$: The base standard deviation of the Gaussian weighting function used for blurring the DoG responses, to allow for some tolerance in the position of the found points (See \autoref{eq:sigmaGaussian}).
	\item $\alpha$: Parameter $\alpha$ that determines the standard deviation of the Gaussian used for blurring the DoG responses. The higher $\alpha$, the more the standard deviation grows for larger concentric circles (See \autoref{eq:sigmaGaussian}).
\end{itemize}

To determine optimal $(\sigma,\rho,\sigma_0,\alpha)$, we split the datasets into a training set and a validation set. We use the training set for finding a good set of $(\sigma,\rho,\sigma_0,\alpha)$ and validate the decision on the entire dataset. We limit the search space of $(\sigma,\rho,\sigma_0,\alpha)$ by results found on specific datasets in previous research, such as the results of Azzopardi \textit{et al}., in which for instance optimal values of $\sigma$ were found for certain resolutions, and this hints us to the region in which the optimal value of $\sigma$ is likely to reside. The procedure for determining the parameter values can then be summarized as follows:

\begin{enumerate}
	\item Split the dataset into an equal sized training set and a validation set, each consisting of $n$ images.
	\item For a search space of $(\sigma_s,\rho_s,\sigma_{0s},\alpha_s)$ which depends on the properties of the dataset, take the filter corresponding to this combination of parameters and filter all $n$ training images with this filter obtaining $n$ response images.
	\item Iterate over a range of threshold $t \in [0,1]$ with intervals of $0.01$. For each threshold $t$, segment all $n$ response images according to Eq. \eqref{eq:thresholding}. This yields $n$ binary images with the vessel segmentation.
	\item Having obtained these $n$ binary images, we can compare all $n$ images with the ground truth vessel segmentation images. Each vessel segmentation for one image $f$ obtained from the filter and threshold combination paired with the ground truth segmentation of image $f$ yields a confusion matrix, as observable in Table \ref{tb:confmat}. We then compute for each of these $n$ confusion matrices the Matthews Correlation Coefficient (MCC), which is directly computable from the corresponding confusion matrix as only quantities from the confusion matrix are used:
	\begin{equation} \label{eq:MCC}
	{\text{MCC}}={\frac  {TP\times TN-FP\times FN}{{\sqrt  {(TP+FP)(TP+FN)(TN+FP)(TN+FN)}}}}
	\end{equation}
	The MCC is particularly suitable for the assessment of binary classifiers, and can even be used if  there is a disbalance between the number of datapoints, in this case pixels, per class. In our use case, the MCC is therefore suitable as vessel segmentation is a binary classification problem with a disbalance between classes, as there are naturally more non-vessel pixels compared to vessel pixels. Having computed the $MCC$ for the responses of all training images thresholded with the current threshold, we average this set of $MCC$ values. After that we consider the next threshold by going to step 4 and compute the new average $MCC$. In the end, the combination of parameters $(\sigma,\rho,\sigma_0,\alpha)$ along with a threshold $t$ that yields the best average $MCC$ on the training set are the parameters that we will use for segmentation of blood vessels in the entire dataset.
\end{enumerate} 
For the best set of filter parameters obtained from the above procedure, we store for each threshold $t$ the corresponding average $TPR$ and $FPR$ over the segmented training images. By plotting the obtained values of $TRP$ against $FPR$, we get a so-called ROC curve. The threshold $t$ that gave the best average $MCC$ value over the segmented training images is the threshold we use for the segmentation of the images in the validation set, and possible future images that are obtained with the same parameters as for the dataset under consideration.

\begin{table}
  \scriptsize
  \begin{center}
    \begin{tabular}{|c|c|c|}
    \hline
     & Classifier: Vessel & Classifier: Non-Vessel\\
    \hline
      GT: Vessel &  TP & FN \\
    \hline
      GT: Non-Vessel & FP & TN \\
     \hline
    \end{tabular}
  \end{center}
  \caption{Confusion matrix: The classifier output (the thresholded filter response) is on the horizontal access. On the vertical access is the Ground Truth, which is the assessment of an expert that manually annotates all the pixels in the image as either vessel or non-vessel.}
  \label{tb:confmat}
\end{table}



\subsection{Datasets}
The retinal images in the IOSTAR dataset were obtained using Scanning Laser Ophtalmoscopy (SLO) \cite{IOSTAR}, \cite{IOSTARLink}. In 2015, the dataset was one of the first publicly available retinal datasets of which the data was obtained using this technique. The resolution of the images are $1024\times 1024$. Based on results presented by Azzopardi \textit{et al}., which show an expected positive correlation between resolution and $\sigma$, we limit the search space to $\sigma=[4.8,4.9,...,5.1]$. We choose $\rho_{max}=20$ as the maximum radius of the concentric circles, and following Azzopardi \textit{et al}. we increment the circles by 2 pixels. Therefore the radii of the circles that are considered are $\rho=[0,2,4,...,20]$. $\sigma_0$ is limited to $\sigma_0=\{1,2,3\}$ and $\alpha$ to $\alpha=\{0.1,0.2,...,0.7\}$.

Once we choose the combination of parameters for the symmetric filter that yields the best average $MCC$ over the images in the training set, we fix these parameters and search for a good combination of parameters for the asymmetric filter. Note that we know in advance that vessel endings are usually thinner, and therefore we search for $\sigma_a$ in a value range that is strictly smaller than the obtained $\sigma_s$. 

\section{Results} \label{sec:results}
\begin{figure}[htb]
  \centering
  \begin{subfigure}[b]{0.22\textwidth}
  	\includegraphics[width=\textwidth]{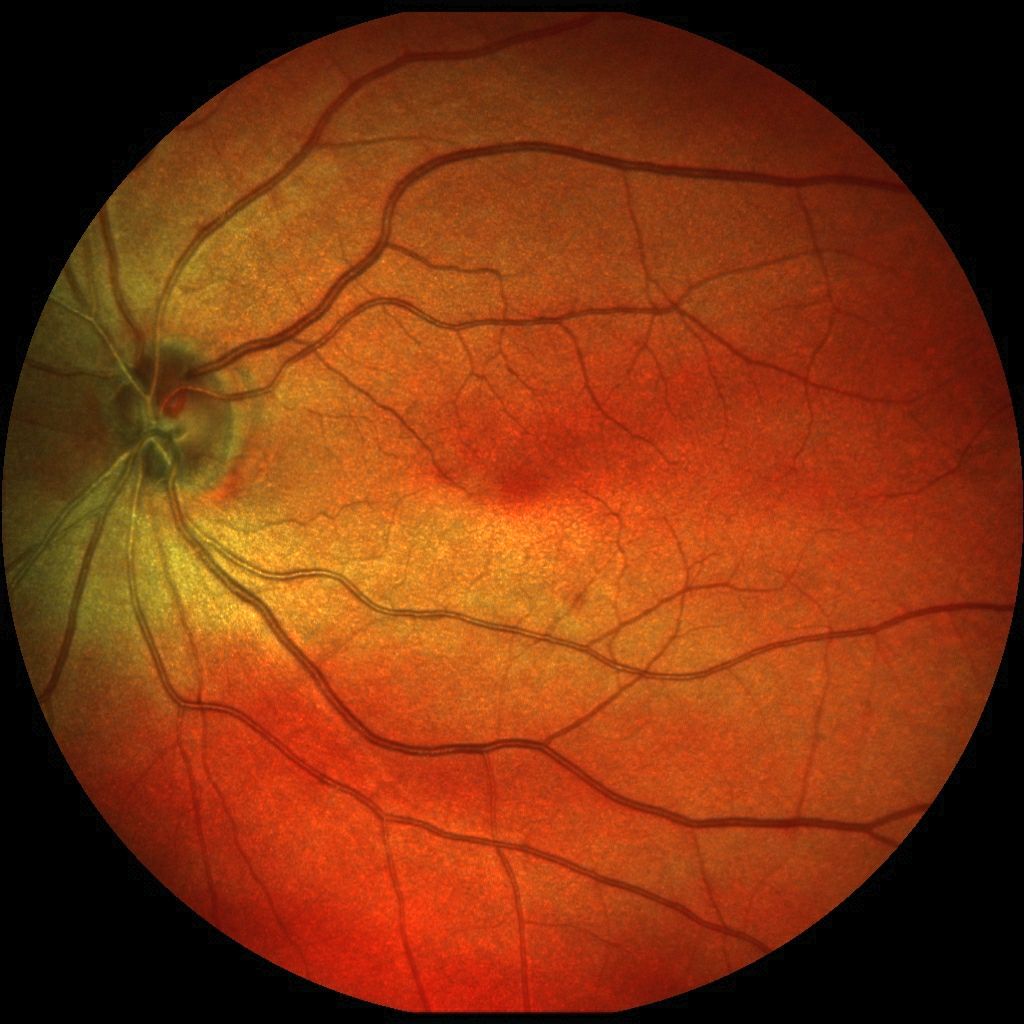}
  	\caption{An IOSTAR retinal fundus image}
  	\label{fig:segmentedGT}
  \end{subfigure}   
  \begin{subfigure}[b]{0.22\textwidth}
  	\includegraphics[width=\textwidth]{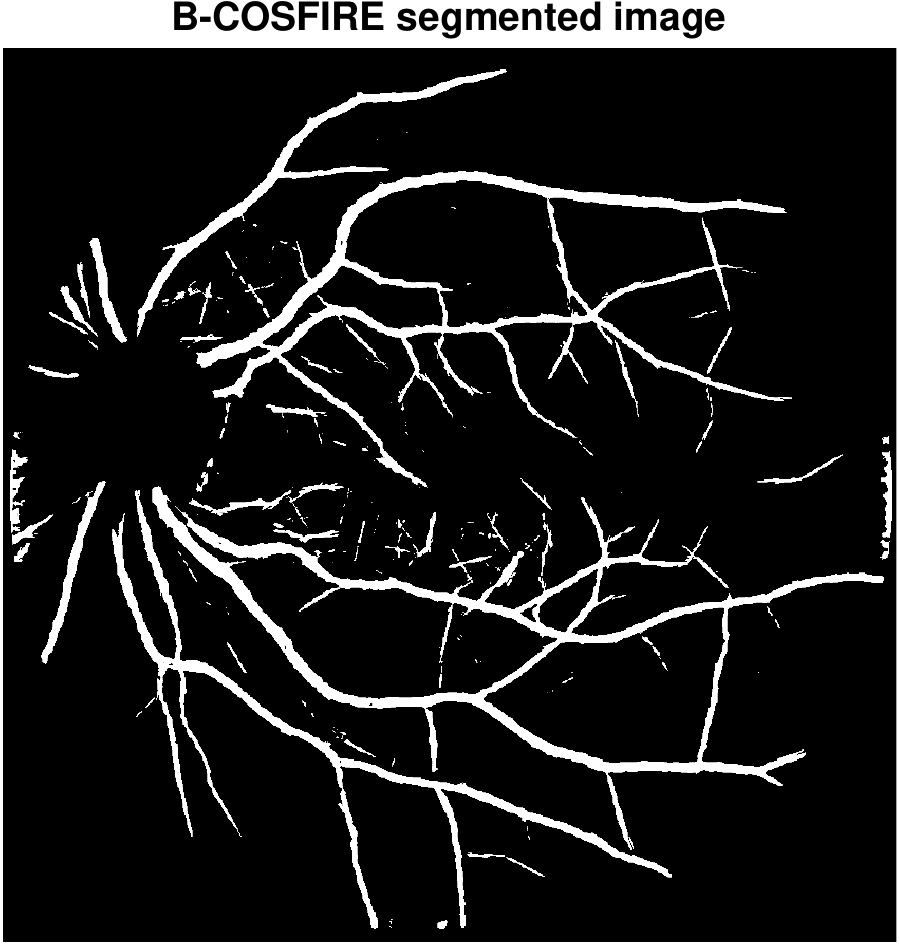}
  	\caption{Segmented with filter}
  	\label{fig:segmented}
  \end{subfigure} 
  \caption{\textit{Left}: An original training image from the IOSTAR dataset. \textit{Right}: The corresponding segmentation using the trained B-COSFIRE filters with parameters specified in \autoref{tb:params1} and threshold $t=35$.}
  \label{fig:segment}
\end{figure}

\begin{figure}[htb]
  \centering
  \includegraphics[width=3.5in]{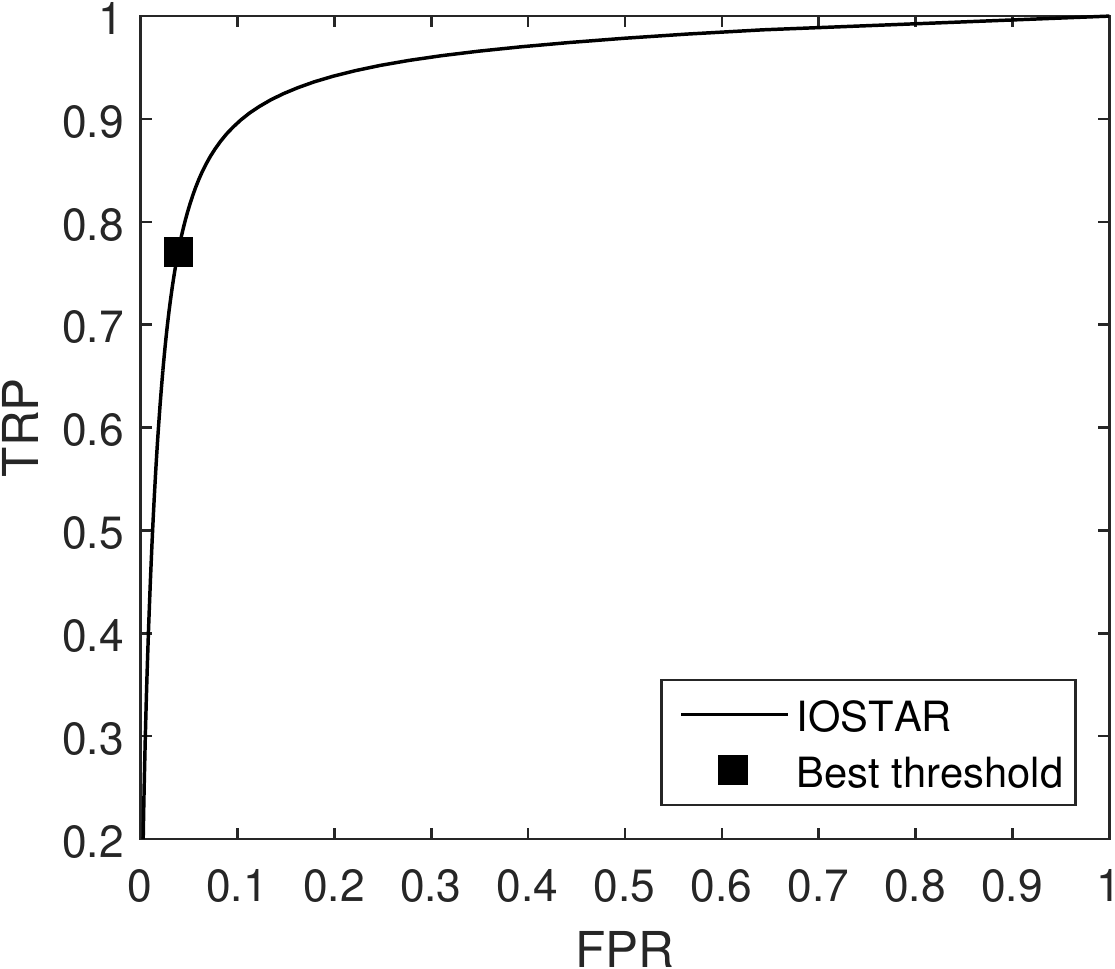}
  \caption{The ROC curve obtained by varying the segmentation threshold from 0 to 255 which yields for each threshold a true positive rate (TPR) and a false positive rate (FPR). Indicated with a square is the TPR and FPR of the threshold $t=35$, which yielded the best average MCC value.}
  \label{fig:ROC}
\end{figure}

In \autoref{fig:ROC} the ROC curve is traced out for the segmentation of the training images in the IOSTAR dataset with the set of parameters in \autoref{tb:params1}, using the manually segmented ground truth images as the correct classification against which the filter segmentation is compared. The threshold with the highest average MCC value is also indicated, which turns out to be a threshold of 35. In \autoref{tb:segAccuracy1} we see the resulting AUC value and also the segmentation accuracy corresponding with the best threshold. In \autoref{fig:segment} we see an original IOSTAR training image alongside the segmented image using the filter parameters in \autoref{tb:params1} and threshold $t=35$. In line with the good performance metrics, we can also see that the resulting segmentation is quite accurate in the sense that the main vessel tree has been segmented correctly and even the tinier vessels are for the most part in the segmentation.
\begin{table}
	\centering
	\begin{tabular}{ccccc}
		\hline
		& $\sigma$ & $\rho_{max}$ & $\sigma_0$ & $\alpha$\\
		\hline
		Symmetric 	& 	4.8	& 20 & 3 & 0.3\\
		Asymmetric 	&	4.4	& 36 & 1 & 0.1\\
	\end{tabular}
	\caption{The parameters used for the symmetric- and the asymmetric filter used for segmentation of blood vessels in the images of the IOSTAR dataset.}
	\label{tb:params1}
\end{table}

\begin{table}
	\centering
	\begin{tabular}{cccccc}
		\hline
		& AUC & MCC & Accuracy & Se & Sp\\
		\hline
		IOSTAR & 0.9519 & 0.6979 & 0.9419 & 0.7705 & 0.9613\\
	\end{tabular}
	\caption{IOSTAR segmentation accuracy with the parameters specified in \autoref{tb:params1}, using the segmentation threshold $t=35$.}
	\label{tb:segAccuracy1}
\end{table}

From an experiment in \cite{BCOSFIRE} on the DRIVE retinal image dataset, it has been established that the parameter $\alpha$ is the most sensitive parameter, i.e., a correct configuration of this parameter is most determinant for the performance of the segmentation. Here we perform a similar experiment for the IOSTAR dataset: From the parameter values as listed in \autoref{tb:params1} for the symmetric filter, we inspect the sensitivity of each parameter at a time by changing the parameter value by steps of 0.1 from the optimal value and keeping the rest of the parameters at their optimal values. We then compare the resulting 30 MCC values with the 30 MCC values obtained for the optimal parameter values in \autoref{tb:params1} by a two-tailed paired t-test with significance level $p=0.05$. We regard the obtained MCC values to be significantly different from the optimal MCC values if $p < 0.05$. The t-values for which this is true are bold-faced in \autoref{tb:ttest}.

\begin{table}
	\centering
	\begin{tabular}{cccc}
		\hline
		 & $\sigma_0$ & $\sigma$ & $\alpha$ \\		
		-0.5 & \textbf{5.0363} & -0.7858 & - \\ 
		-0.4 & \textbf{3.7276} & -0.5248 & - \\ 
		-0.3 & \textbf{2.4602} & -0.3723 & -0.9215 \\ 
		-0.2 & 1.3496 & -0.1580 & 1.7180 \\
		-0.1 & 0.5614 & -0.0862 & \textbf{5.8397} \\
		 0	 & - & - & - \\
		+0.1 & -0.6672 & -0.4170 & \textbf{-2.4933} \\
		+0.2 & -1.1272 & -0.9986 & \textbf{-3.6321} \\
		+0.3 & -1.5860 & -1.9217 & \textbf{-4.3719} \\
		+0.4 & -1.9512 & \textbf{-2.5296} & \textbf{-4.9806} \\
		+0.5 & \textbf{-2.2660} & \textbf{-3.1036} & \textbf{-5.4493} \\
	\end{tabular}
	\caption{An experiment to study the sensitivity of the parameters $\sigma_0$, $\sigma$ and $\alpha$ for a symmetric B-COSFIRE filter on the 30 images of the IOSTAR dataset. Each row compares the 30 MCC values for the optimal parameters with the 30 MCC values for the parameter values for the corresponding row. Each row indicates the t-statistic after changing the parameter value by the offset in the first column from the optimal parameter value and keeping the remaining two parameters at their optimal values. Bold-faced are the t-values for which the null-hypothesis is rejected with significance level $p < 0.05$.}
	\label{tb:ttest}
\end{table}

As can be see in \autoref{tb:ttest}, the parameters have a different level of sensitivity. From our experiments, the parameter $\alpha$ turns out to be the most sensitive parameter. By a small change the t-statistic changes considerably and we found that even a change of 0.1 yields a significantly different segmentation performance. We observe that the parameters $\sigma_0$ or $\sigma$ are less sensitive, as the change in t-value is smaller compared to the case of $\alpha$. However, we notice that the $\sigma$ parameter is more sensitive when it is increased from $\sigma=4.8$ than if it is decreased. The reverse is true for $\sigma_0$: It is less sensitive when increased from $\sigma_0=3$ and more sensitive when decreased. The result that $\alpha$ is the most sensitive parameter is in line with the result in \cite{BCOSFIRE}.

\section{Summary and discussion}
In this contribution we discussed various methods for the automatic segmentation of vessels in retinal fundus images. In this section we discuss how they perform compared to the B-COSFIRE method and our insights about the B-COSFIRE method.

\subsection{Deep Neural Networks}
Given the results of Liskowski \textit{et al}., their neural network aproach performs better than B-COSFIRE in general. In all cases, the neural network achieves a higher AUC value than B-COSFIRE. However, in terms of accuracy and specificity, B-COSFIRE does outperform the balanced configuration of the network on the DRIVE and STARE dataset. For the CHASE dataset, the neural network outperforms B-COSFIRE.

However, in terms of training, the B-COSFIRE method is cheaper, as Liskowski \textit{et al}. report that training the network can take up to 8 hours on a single GPU. However, training times are not interesting for the end user, as one usually does not need to train the classifier for each new image. To classify an image, the neural network needs 92 seconds using a high end GPU. So, for end users with low end hardware and a limited budget, the B-COSFIRE method could be a better choice in terms of time, because the B-COSFIRE method only needs 10 seconds on a 2 GHz CPU. However, if the hardware or budget is available, the neural network is suitable, because we consider 92 seconds still an acceptable time, given the experiment results.

\subsection{Ensemble classification-based approach}
Compared to the B-COSFIRE method, the method of Fraz \textit{et al}. is more accurate when tested on the three datasets (DRIVE, STARE and CHASE), however, the B-COSFIRE method has a higher sensitivity or true positive rate (TPR), i.e. B-COSFIRE is more able to detect \textit{vessel} pixels. The method of Fraz \textit{et al}. performs better than B-COSFIRE, but the B-COSFIRE is faster. B-COSFIRE only needs 10 seconds to process an image from the DRIVE or STARE datasets whereas the ensemble classification method needs 2 minutes.

The method of Fraz \textit{et al}. performs slightly better than B-COSFIRE, but requires expensive training with retinal fundus images. The parameters of the B-COSFIRE method must either be tweaked manually by hand or be determined using a training set of the images. Based on experiments with the intuitive parameters of B-COSFIRE, manually tweaking the parameters quickly yields a good performance which is a major advantage of the B-COSFIRE method.

\subsection{Fully Connected Conditional Random Field Model}
Orlando \textit{et al}. did not provide AUC and accuracy metrics, therefore, we only use the sensitivity and specificity to compare the Fully Connected Conditional Random Field method with B-COSFIRE. As it turns out, there is no winner. For the DRIVE dataset, Orlando \textit{et al}. achieve higher sensitivity, but lower specificity. For the STARE dataset, it is the other way around and for the CHASE dataset, Orlando \textit{et al}. achieve also less sensitivity and more specificity.

\subsection{B-COSFIRE}
Furthermore, we have studied the inner workings of the B-COSFIRE method and subsequently applied the method for a set of parameters to the recent IOSTAR dataset. We have seen that B-COSFIRE could achieve an accurate and fast segmentation on this dataset.

B-COSFIRE has turned out to be a fast trainable filter-based method that can definitely compete with much more complex methods that are based on machine learning methods like neural networks, that take a considerable amount of time to train. Given the complexity of those methods and B-COSFIRE's potential to compete well and its intuitive parameters, B-COSFIRE could even be the preferred method to use, especially when having less computational power at hand.

\subsection{Other fields and future work}
Segmentation methods like the ones we have discussed in this paper are not solely useful for blood vessel segmentation. In fact, one could use these methods to segment rivers from satellite images, segment the nerves of a leaf, to find cracks in concrete structures, in fact these methods could provide useful in all applications in which elongated structures need to be segmented.

For future work, any of these methods could be extended to not only extract the blood vessel tree but to also detect obstacles that could impede the blood flow. An additional benefit of B-COSFIRE is that its selectivity is not pre-programmed into the method itself but it is achieved by prototype patterns on which the filter is automatically configured. This gives the possibility to extend the filter's selectivity to all kinds of more complex structures which gives rise to applications other than vessel segmentation alone.


\bibliographystyle{ieeetr}
\bibliography{bloodVessels}

\begin{thebibliography}{10}

\bibitem{BCOSFIRE}
G.~Azzopardi, N.~Strisciuglio, M.~Vento, and N.~Petkov, ``Trainable cosfire
  filters for vessel delineation with application to retinal images,'' {\em
  Medical image analysis}, vol.~19, pp.~46--57, 1 2015.

\bibitem{Fraz}
M.~M. Fraz, P.~Remagnino, A.~Hoppe, B.~Uyyanonvara, A.~R. Rudnicka, C.~G. Owen,
  and S.~A. Barman, ``An ensemble classification-based approach applied to
  retinal blood vessel segmentation,'' {\em IEEE Transactions on Biomedical
  Engineering}, vol.~59, pp.~2538--2548, Sept 2012.

\bibitem{Krawiec}
P.~Liskowski and K.~Krawiec, ``Segmenting retinal blood vessels with deep
  neural networks,'' {\em IEEE Transactions on Medical Imaging}, vol.~35,
  pp.~2369--2380, Nov 2016.

\bibitem{Prokofyeva}
J.~I. Orlando, E.~Prokofyeva, and M.~B. Blaschko, ``A discriminatively trained
  fully connected conditional random field model for blood vessel segmentation
  in fundus images,'' {\em IEEE Transactions on Biomedical Engineering},
  vol.~64, pp.~16--27, Jan 2017.

\bibitem{IOSTAR}
J.~Zhang, B.~Dashtbozorg, E.~Bekkers, J.~P.~W. Pluim, R.~Duits, and B.~M. ter
  Haar~Romeny, ``Robust retinal vessel segmentation via locally adaptive
  derivative frames in orientation scores,'' {\em IEEE Transactions on Medical
  Imaging}, vol.~35, pp.~2631--2644, Dec 2016.

\bibitem{COSFIRE}
G.~Azzopardi and N.~Petkov, ``Trainable cosfire filters for keypoint detection
  and pattern recognition,'' {\em Ieee transactions on pattern analysis and
  machine intelligence}, vol.~35, pp.~490--503, 2 2013.
\newblock Relation: http://www.rug.nl/research/jbi/ Rights: University of
  Groningen, Johann Bernoulli Institute for Mathematics and Computer Science.

\bibitem{staal:2004-855}
J.~Staal, M.~Abramoff, M.~Niemeijer, M.~Viergever, and B.~van Ginneken,
  ``{Ridge based vessel segmentation in color images of the retina},'' {\em
  {IEEE Transactions on Medical Imaging}}, vol.~23, no.~4, pp.~501--509, 2004.

\bibitem{1677726}
A.~M. Mendonca and A.~Campilho, ``Segmentation of retinal blood vessels by
  combining the detection of centerlines and morphological reconstruction,''
  {\em IEEE Transactions on Medical Imaging}, vol.~25, pp.~1200--1213, Sept
  2006.

\bibitem{1677727}
J.~V.~B. Soares, J.~J.~G. Leandro, R.~M. Cesar, H.~F. Jelinek, and M.~J. Cree,
  ``Retinal vessel segmentation using the 2-d gabor wavelet and supervised
  classification,'' {\em IEEE Transactions on Medical Imaging}, vol.~25,
  pp.~1214--1222, Sept 2006.

\bibitem{4336179}
E.~Ricci and R.~Perfetti, ``Retinal blood vessel segmentation using line
  operators and support vector classification,'' {\em IEEE Transactions on
  Medical Imaging}, vol.~26, pp.~1357--1365, Oct 2007.

\bibitem{niemeijer:2004-866}
M.~Niemeijer, J.~Staal, B.~van Ginneken, M.~Loog, and M.~Abramoff,
  ``{Comparative study of retinal vessel segmentation methods on a new publicly
  available database},'' in {\em {SPIE Medical Imaging}} (J.~M. Fitzpatrick and
  M.~Sonka, eds.), vol.~5370, pp.~648--656, {SPIE}, {SPIE}, 2004.

\bibitem{Fraz2012407}
M.~Fraz, P.~Remagnino, A.~Hoppe, B.~Uyyanonvara, A.~Rudnicka, C.~Owen, and
  S.~Barman, ``Blood vessel segmentation methodologies in retinal images – a
  survey,'' {\em Computer Methods and Programs in Biomedicine}, vol.~108,
  no.~1, pp.~407 -- 433, 2012.

\bibitem{Srivastava:2014:DSW:2627435.2670313}
N.~Srivastava, G.~Hinton, A.~Krizhevsky, I.~Sutskever, and R.~Salakhutdinov,
  ``Dropout: A simple way to prevent neural networks from overfitting,'' {\em
  J. Mach. Learn. Res.}, vol.~15, pp.~1929--1958, Jan. 2014.

\bibitem{DRIVELink}
``{DRIVE Dataset}.'' \url{http://www.isi.uu.nl/Research/Databases/DRIVE/},
  2004.
\newblock [Online; accessed March 5th, 2017].

\bibitem{Hoover00locatingblood}
A.~Hoover, V.~Kouznetsova, and M.~Goldbaum, ``Locating blood vessels in retinal
  images by piecewise threshold probing of a matched filter response,'' {\em
  IEEE Transactions on Medical Imaging}, vol.~19, pp.~203--210, 2000.

\bibitem{STARELink}
``{STARE Dataset}.'' \url{http://cecas.clemson.edu/~ahoover/stare/}, 2004.
\newblock [Online; accessed March 5th, 2017].

\bibitem{CHASELink}
``{CHASE Dataset}.'' \url{https://blogs.kingston.ac.uk/retinal/chasedb1/}.
\newblock [Online; accessed March 5th, 2017].

\bibitem{IOSTARLink}
``{IOSTAR Dataset}.'' \url{http://www.retinacheck.org/}, 2015.
\newblock [Online; accessed 27-Febr-2017].

\end{thebibliography}
\end{document}